\documentclass{WileyMSP-template}

\usepackage{amsmath}

\usepackage[version=3]{mhchem} 
\usepackage{graphicx}
\usepackage{dcolumn}
\usepackage{bm}
\usepackage{siunitx}
\usepackage{amsmath}
\usepackage[dvipsnames]{xcolor}
\usepackage[version=3]{mhchem} 
\usepackage{siunitx}  
\usepackage{multirow} 
\usepackage{comment}
\usepackage{array}
\usepackage{textpos}
\usepackage{rotating} 
\usepackage{esvect}

\usepackage{graphicx}
\usepackage{dcolumn}
\usepackage{bm}
\usepackage{float}

\usepackage[utf8]{inputenc}
\usepackage{float}
\usepackage[T1]{fontenc}
\usepackage{mathptmx}
\usepackage{siunitx}  
\usepackage{amsmath}
\usepackage[hidelinks]{hyperref}
\usepackage[all]{hypcap}
\usepackage[noabbrev, capitalise]{cleveref}
\usepackage{amssymb}

\usepackage[numbers]{natbib}

\begin{document}

\pagestyle{fancy}
\rhead{\includegraphics[width=2.5cm]{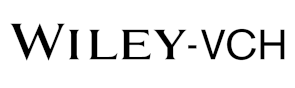}}

\title{Field-driven nonlinear metasurface: self-adaptive transition between high-selectivity transmission and broadband shielding}

\maketitle

\author{Yuan Xu$^{1,3}$}
\author{Hanqing Liu$^{*}$$^{1,3}$}
\author{Shangjing Xi$^{1}$}
\author{Xiaodi Zhang$^{1}$}
\author{Yanqing Cheng$^{1}$}
\author{Wenye Ji$^{*}$$^{2}$}
\author{Peiguo Liu$^{*}$$^{1}$}

\begin{affiliations}

1. College of Electronic Science and Technology, National University of Defense Technology, Changsha 410073, Hunan Province, China.\\
2. Air and Missile Defense College, Air Force Engineering University, Xi'an 710051, Shaanxi Province, China.\\
3. These authors contributed equally to this work.\\ 
Email Address: liuhanqing@nudt.edu.cn; jiwenyewave@163.com; pg731@126.com
\end{affiliations}


\begin{abstract}

{The escalating complexity of electromagnetic (EM) environment is posing a significant challenge to the reliability of modern electronic information systems. To address the need for the spatial EM safety of electronic devices, we present a field-driven nonlinear metasurface (NMS) that enables effective protection against out-of-band interference and in-band high-intensity radiation. By constructing a reconfigurable hybrid coupling topology and mapping it to the metasurface geometry, the proposed NMS achieves a self-adaptive transition between its transmission and shielding mode depending on the incident power. The experimental results are in good agreement with theoretical analysis and full-wave simulation. We obtain a highly selective passband with roll-off rate larger than 20.6~\si{dB/GHz} and a broadband shielding with shielding effectiveness exceeding 23.6~\si{dB} and 60~$\%$ bandwidth, demonstrating a significantly enhanced performance relative to the literature. Our findings establish a promising route toward comprehensive EM protection on radio frequency front-end systems. }

\end{abstract}

\section{Introduction} 

\begin{figure}[H]
  \includegraphics[width=\linewidth]{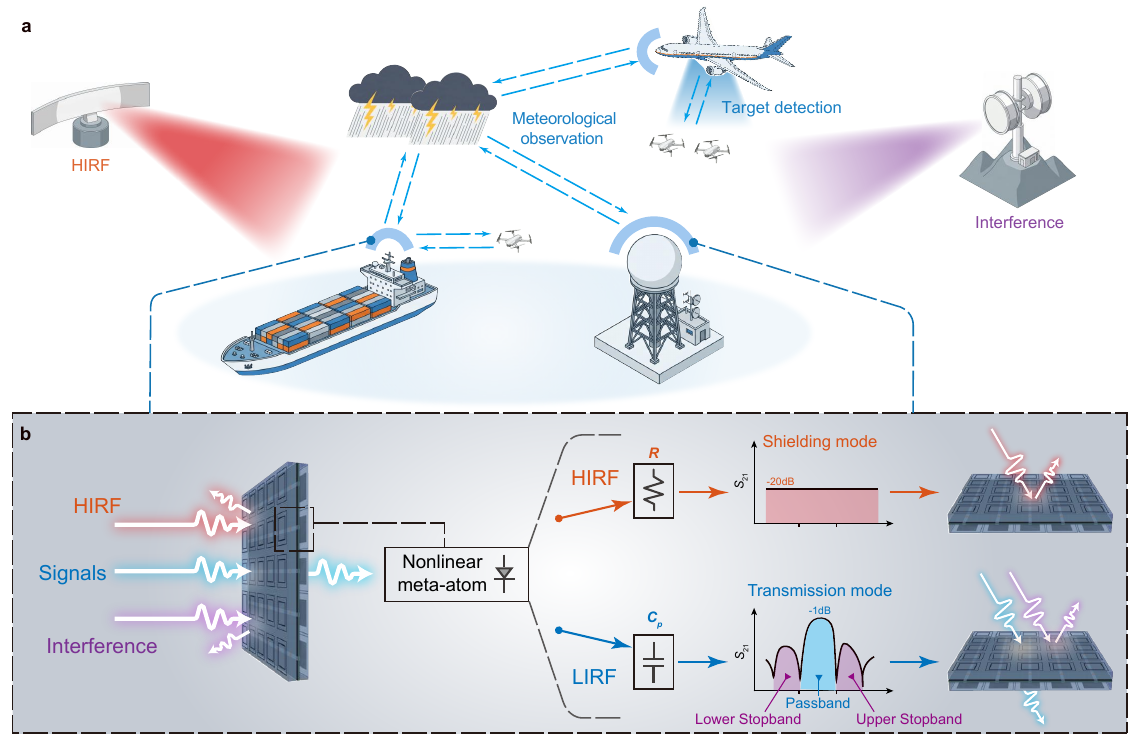}
  \caption{Conceptual schematic of the proposed field-driven self-adaptive nonlinear metasurface (NMS). \textbf{a}. Under HIRF and interference, NMS can be used in various types of electronic information platforms, providing comprehensive support for sensitive radar functions such as target detection and meteorological observation. \textbf{b}. Intensity-dependent operating mechanism of the proposed metasurface. Under low-intensity radiated fields (LIRF), the metasurface exhibits impedance matching with free space only within the target passband, enabling out-of-band interference suppression and highly selective low-loss transmission. Under high-intensity radiated fields (HIRF), the embedded diodes switch to the low-resistance state, short-circuiting the coupling topology and inducing broadband impedance mismatch, thereby establishing broadband shielding.}
  \label{fig:concept}
\end{figure}

The development of wide-bandgap semiconductor materials and advanced packaging technologies has equipped RF electronic devices with high radiated power, broad bandwidth, and enhanced sensitivity \cite{woo2024wide, pengelly2012review, lau2022recent, yi2025through, nam2021high}. Although these attributes facilitate high-throughput data transmission for 5G/6G communications and improve the high-resolution detection capabilities of radar systems \cite{hong2017multibeam}, they simultaneously raise the ambient electromagnetic (EM) noise floor in intricate environments, making the electronic information equipment increasingly susceptible to electromagnetic interference (EMI) and high-intensity radiated fields (HIRF) threats. Therefore, safeguarding core electronic components against the complicated radiation environment has emerged as a pressing challenge in recent years \cite{ESSEleftheriades2014,ESSLi2020, zheng2019radar}. The effect of EMI and HIRF on RF equipment is not the same: EMI is typically transient, causing reversible degradation of receiver sensitivity and signal-to-noise ratio \cite{chen2021high, ESSvan2015band}, while HIRF carries substantial energy, posing a severe risk of permanent physical burnout to internal RF components \cite{ESSyang2013novel, ESS11322717}. As illustrated in \textbf{Figure~\ref{fig:concept}a}, X-band radar platforms, which can precisely detect the low-power target signals from meteorological formations \cite{antonini2017implementation,matrosov2002x} and unmanned aerial vehicles (UAVs) \cite{horstmann2021coherent}, are sensitive to the EMI from fixed wireless systems and the HIRF emitted by adjacent high-power transmitters \cite{li2025study}. This leads to saturation and potential breakdown in the receiver channel of the radar, hindering the capacity to continuously track weak targets and thereby reducing the overall situational awareness of the system.

Nonlinear metasurface (NMS) has emerged as a versatile platform for diverse manipulation of EM waves, including resonance tunability \cite{yu2022electrically,jangid2024spectral},  nonreciprocal propagation \cite{li2026tunable,cao2025broadband,guo2019nonreciprocal}, wavefront steering \cite{wang2018nonlinear,rong2023beam}, intensity-dependent spatial filtering \cite{ESSDEV_4_tian2025energy,ESSDEV_5, liu2023electromagnetic}, wave mixing \cite{ke2021linear} and localization \cite{luo2016electrically,huang2020planar}. The subwavelength meta-atoms, either electrically active or field-driven, provide NMS with precise customization and substantial amplification of nonlinear effects \cite{luo2022high}. A representative example is the energy selective surface (ESS), which is composed of nonlinear switches and resonant structures, can achieve power-dependent transmission in accordance with the incident power \cite{ hu2023design}. Previous studies on such designs have mainly focused on the in-band characteristics and were capable of achieving in-band shielding against HIRF \cite{hu2022design}. However, the optimization for out-of-band performance is usually ignored, especially for a high selectivity that can significantly suppress out-of-band interference. The roll-off rate, defined as the slope at the edge of the passband where the signal attenuation varies with frequency \cite{xiong2024miniaturized, kanth2019design}, determines the out-of-band interference suppression of the metasurface at its adjacent stopbands. Although conventional filtering-type metasurfaces with high roll-off rates have been widely studied \cite{hong2021highly, zhao2019bandpass}, an NMS capable of effectively safeguarding against both HIRF and out-of-band interference remains a challenge, as it requires a complicated design with respect to coupling topologies and structural mapping.

In this work, we propose a field-driven NMS that enables self-adaptive transitions between high-selectivity transmission and broadband shielding. As depicted in \textbf{Figure~\ref{fig:concept}b}, due to a change in electrical properties ($R \longleftrightarrow C_p$) of the meta-atom, the proposed structure exhibits an adaptive response depending on the incident power: under low-intensity radiated fields (LIRF), the NMS maintains low-loss transmission with an insertion loss (IL) below 1~\si{dB} in its passband and simultaneously suppresses the interference in its lower and upper stopbands; while under HIRF, the NMS transitions to high shielding with a shielding effectiveness (SE) exceeding 20~\si{dB} across a broad frequency range. We built a hybrid coupling network topology consisting of three resonators to realize high selectivity with steep roll-off rates, and then mapped the topology onto a five-layer geometrical structure. The field distributions of the proposed NMS reveal distinct mechanisms for the transmission zeros (TZs), resonant passband tunneling, and shielding-mode impedance collapse. The measured EM characteristics under both LIRF and HIRF agree well with the theoretical analysis and full-wave simulation.

\section{Results and Discussion}

\subsection{Achieving the high-performance EM protection via equivalent circuit analysis}

\begin{figure}[H]
  \includegraphics[width=\linewidth]{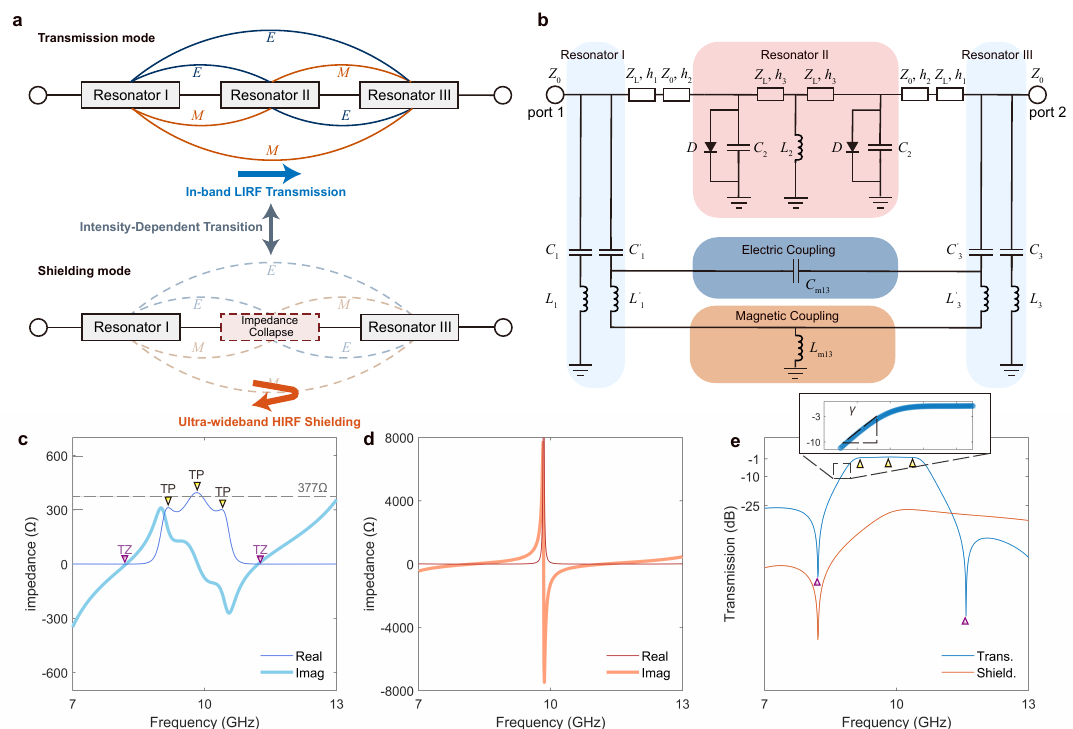}
  \caption{Field-driven hybrid-coupling mechanism and the corresponding equivalent-circuit response of the proposed NMS. \textbf{a}. Field-driven hybrid coupling network topology. \textbf{b}. Equivalent circuit model (ECM) corresponding to the topology. Real and imaginary parts of the equivalent surface impedance $Z_{\text{m}}$ in \textbf{c}. highly matched transmission mode and \textbf{d}. severely mismatched shielding mode. \textbf{e}. Calculated $S$ parameters in the transmission and shielding modes, confirming the unification of high-selectivity transmission and broadband shielding.}
  \label{fig:theory_model}
\end{figure}

We start by building a field-driven hybrid coupling network topology to realize the adaptive protection of NMS. As illustrated in \textbf{Figure~\ref{fig:theory_model}a}, the proposed topology consists of three mutually coupled sub-wavelength resonators. Under LIRF, attributed to the electrical ($E$) and magnetic ($M$) coupling effects among all resonators, a highly-selective transparent window forms in the operating band, enabling the incident waves to transmit with low loss (transmission mode); under HIRF, the high-density field will instantaneously trigger resonator II to a low-impedance shunt path connected with the ground, resulting in effective shielding against the incident waves (shielding mode). We then establish an equivalent circuit model (ECM) that provides physical insights into the underlying mechanisms governing the NMS, as shown in \textbf{Figure~\ref{fig:theory_model}b}. Here, resonators I and III are constructed using a dual-branch series-LC topology, which offers the necessary topological degrees of freedom to achieve the desired high-selectivity quasi-elliptic response. The cross-coupling between resonators I and III is realized through $C_{m13}$ and $L_{m13}$. To implement the field-driven adaptive transition, the central resonator II is configured as a parallel $LC$ circuit ($L_2C_2$), where two semiconductor diodes ($D$) are loaded to dynamically reconfigure the local impedance. As a switch, the diode does not respond to normal signals and remains as a static capacitance $C_{\text{off}}$. In the OFF state, this hybrid coupling network creates the essential bypass pathway required for multi-channel destructive interference, as analyzed in SI section S1 of the Supporting Information. Conversely, once the induced voltage reaches the turn-on voltage, the diode transfers to a resistance $R_{\text{on}}$, allowing electric charges to flow into the ground. The employed elementary LC units also ensure a direct mapping from ECM elements to planar metallic meta-atoms when designing the metasurface.

Using the equivalent medium theory, we calculate $S_{21}$ of the proposed metasurface given by $|S_{21}|^2 = 1 - |\Gamma|^2$, where the reflection $\Gamma$ is equal to $\frac{Z_{\text{m}}-Z_0}{Z_{\text{m}}+Z_0}$, $Z_{\text{m}}$ is the surface impedance of the metasurface, and $Z_0 \approx 377~\Omega$ is the intrinsic wave impedance in free space. Given the values of all $LC$ components as listed in Table~S1, extracted via Advanced Design System (ADS) optimization, we obtain the results of $Z_{\text{m}}$ versus frequency for the built circuit via simulation, as plotted in \textbf{Figure~\ref{fig:theory_model}c} and \textbf{\ref{fig:theory_model}d}. In the transmission mode, three resonant modes appear in the impedance spectrum, corresponding to three transmission poles (TPs) of the circuit model. The values of $\text{Re}(Z_{\text{m}})$ are close to $Z_0$ around the TPs, indicating good impedance matching between the metasurface and free space. The excited TPs merge together to form a transmission window from 9.1 to 10.5~GHz. As shown in \textbf{Figure~\ref{fig:theory_model}e} (blue line), the signal in this frequency range can pass through the metasurface with an IL below $1$ dB. On both sides beyond the passband, two TZs in the $S_{21}$ response can be found at around 8.2 and 11.6~GHz, respectively. These result in a significant decrease of $S_{21}$ in the out-of-band region that is capable of both a quasi-elliptic filtering response and effective anti-interference. Analysis in SI section S1 of the Supporting Information shows that the cross-coupling predominantly influences the upper-sideband TZ characteristics. We define the slope $\gamma = |\Delta S_{21} / \Delta f|$ as the roll-off rate of the transmission curve, where $\Delta S_{21}$ denotes a reduction from $-3$~dB to $-10$~dB, as illustrated in the inset of Figure~\ref{fig:theory_model}e. The extracted roll-off rates of 33.3~dB/GHz at the lower skirt and 35~dB/GHz at the upper skirt confirm the high-selectivity filtering mechanism. In the shielding mode, $\text{Re}(Z_{\text{m}})$ is much higher than $Z_0$ and $\text{Imag}(Z_{\text{m}})$ exhibits a strong reactive fluctuation. The resulting impedance mismatch yields broadband shielding with a SE exceeding 27.1~dB across 7--13~GHz.

\subsection{Designing nonlinear electromagnetic metasurface by full-wave simulation}

\begin{figure}[H]
  \includegraphics[width=\linewidth]{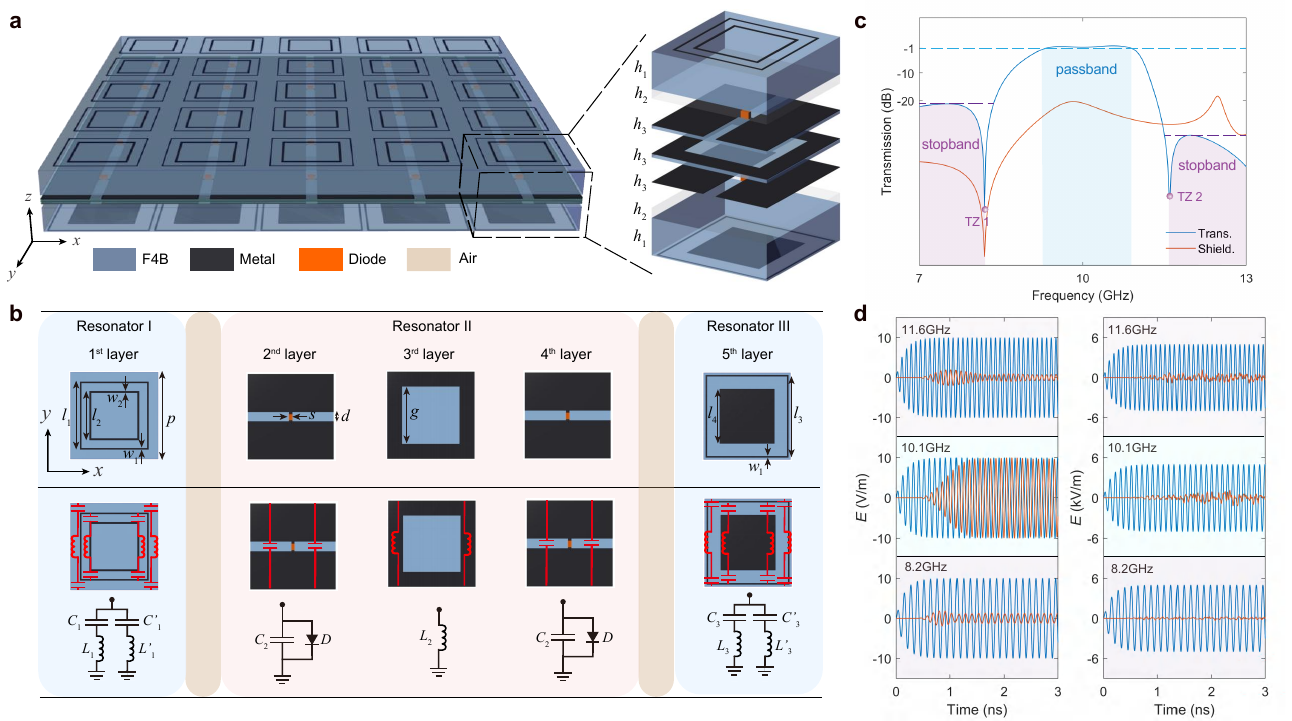}
  \caption{Structural realization and full-wave characterization of the proposed NMS. \textbf{a}. Schematic of the NMS and the corresponding 3D exploded view of the sub-wavelength five-layer meta-atom. \textbf{b}. Layer-by-layer geometry and mapping from the geometry to the ECM topology. \textbf{c}. Full-wave simulated transmission coefficient $S_{21}$ in the transmission and shielding modes marked with passband, stopbands and transmission zeros (TZs). \textbf{d}. Simulated transient characteristics under LIRF ($10~\text{V/m}$) and HIRF ($5~\text{kV/m}$) excitations. Under LIRF, the transmitted wave at $10.1~\text{GHz}$ maintains a stable amplitude, while the response is suppressed at the TZs; under HIRF, the output waveform is clamped, indicating the nonlinear transition to the shielding mode.}
  \label{fig:3D_model}
\end{figure}

We now focus on the construction of NMS that enables a quasi-elliptic response and adaptive protection following the built topology. As shown in the schematic configuration in \textbf{Figure~\ref{fig:3D_model}a}, the proposed NMS is made of three resonators (I to III) on top of dielectric substrates. A five-layer structure is thus adopted as the periodic unit of NMS, where the mapping relation from geometry to circuit topology can be found in \textbf{Figure~\ref{fig:3D_model}b}. For resonator I, the outer and inner metal loops on the 1$^{\text{st}}$ layer are adopted to denote the series $L_1C_1$ and $L'_1C'_1$, respectively; for resonator II, two square capacitive patches featuring a central straight-opening seam are utilized to form $C_2$ on the 2$^{\text{nd}}$ and 4$^{\text{th}}$ layers, and a square aperture slot is used to form $L_2$ on the 3$^{\text{rd}}$ layer; for resonator III, the outer metal loop and inner patch on the 5$^{\text{th}}$ layer are adopted to form the series $L_3C_3$ and $L'_3C'_3$. To enable the adaptive EM protection, a PIN diode (MA4AGP907) is individually integrated in the center of straight-opening seams on the 2$^{\text{nd}}$ and 4$^{\text{th}}$ layers. Once the voltage difference across the diode reaches its turn-on threshold, $C_2$ is short-circuited and the electric charges directly flow into the ground.

Although the mapping illustrates the physical assembly of each layer, cascading these layers into a highly integrated structure needs to introduce the complex near-field interactions. The relationship between the metal-structure parameters and the lumped-element values is analyzed using the analytical impedance models in SI section S2, from which the initial structural parameters of each metallic layer are estimated. Accordingly, we optimize the precise dimensions of the metasurface via full-wave EM simulation in CST Studio Suite. Periodic boundary conditions and Floquet port excitations are set for the unit. The geometrical parameters in mm-scale are fixed as follows: $p = 10.16$ ($\approx \lambda/3$ at $10 \text{ GHz}$), $l_1 = 7.9$, $l_2 = 5.7$, $l_3 = 9.66$, $l_4 = 6.35$, $w_1 = 0.15$, $w_2 = 0.2$, $g = 6.7$, $s = 0.35$, and $d = 1.1$. Considering the impedance matching between the metasurface and free space, the layer thicknesses are $h_1 = 4$, $h_2 = 1$, and $h_3 = 0.254$. The diode MA4AGP907 has a conductive resistance $R_{\text{ON}} = 4.2~\Omega$ and a cutoff capacitance $C_{\text{OFF}} = 30~fF$ from the manufacturer datasheet. By substituting the given values above into the simulation, we obtain the $S_{21}$ curves versus frequency for the proposed NMS, as shown in \textbf{Figure~\ref{fig:3D_model}c}, which are in good agreement with ADS calculation results. In transmission mode, a transparent window (passband) can be observed from 9.3 to 10.9~\si{GHz} with IL $\leq$ 1~\si{dB}, which is tightly confined by the sharp transition skirts dominated by the TZs located at 8.2 and 11.6~\si{GHz}. The roll-off rates reach 17.5 and 41.2~dB/GHz at the lower and upper transition skirts, respectively. The out-of-band suppression reaches 21.2~dB in the lower stopband and 32.6~\si{dB} in the upper stopband. In the shielding mode, the impedance collapse yields broadband shielding with an SE exceeding 17.6~\si{dB} across the 7 to 13~\si{GHz} band, while the SE within the original passband peaks at around 26.8~\si{dB}. 

We further investigate the transient response of the proposed metasurface using the field-circuit co-simulator in CST Studio Suite (see \textbf{Methods} for detailed configurations). Here, the time-domain input and output signals are obtained at the simulated passband center (10.1~\si{GHz}) and at the two simulated TZ frequencies (8.2 and 11.6~\si{GHz}), as plotted in \textbf{Figure~\ref{fig:3D_model}d}. Under LIRF with an incident field of 10~\si{V/m}, the NMS operates in the high-selectivity transmission mode. The initial zero-amplitude interval in the transmitted waveform is ascribed to the propagation delay between the excitation port and the receiving port, while the slight initial envelope deformation in the transmitted wave is caused by the energy build-up process of the resonant metasurface prior to reaching its steady state. On the other hand, under HIRF with an incident field of 5~\si{kV/m}, the output waveforms are clamped across the spectrum. The observed distortion in the transmitted waveform can be attributed to the higher-order harmonics produced by the integrated semiconductor functioning in its nonlinear regime. These transient profiles confirm that the field-driven broadband impedance collapse severs the spatial coupling topology, ensuring a self-adaptive nonlinear transition toward the shielding mode.

\begin{figure}
\centering
  \includegraphics[width=0.8\linewidth]{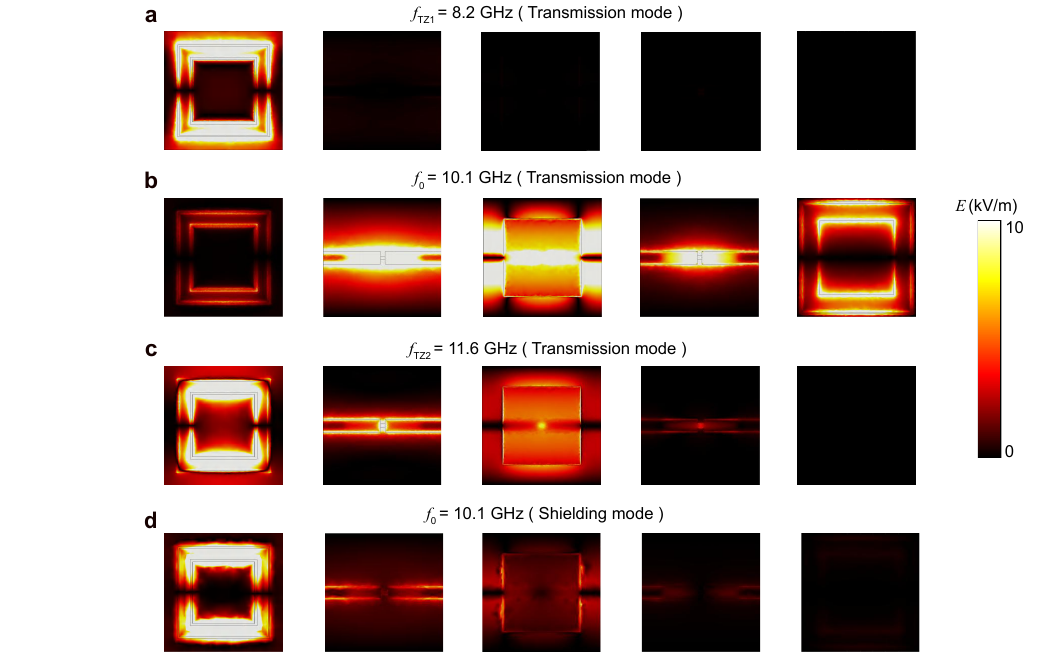}
  \caption{Layer-wise spatial electric-field ($E$) distributions illustrating the underlying field-modulation mechanisms of the proposed NMS. Transmission mode: \textbf{a}. at $f_{\text{TZ1}}$ = 8.2~\si{GHz}, the incident energy is confined to the 1$^{\text{st}}$ layer due to the intrinsic series resonance of the primary cascaded path; \textbf{b}. at $f_0$ = 10.1~\si{GHz}, a resonant tunneling channel is established, allowing the field to penetrate all stacked layers; \textbf{c}. at $f_{\text{TZ2}}$ = 11.6~\si{GHz}, multi-path phase cancellation leads to the suppression of transmission, leaving the 5$^{\text{th}}$ layer unexcited. Shielding mode: \textbf{d}. at $f_0$ = 10.1~\si{GHz}, the impedance collapse restricts the propagating energy to the 1$^{\text{st}}$ layer by short-circuiting the previously transparent window.}
  \label{fig:field_simu}
\end{figure}

The layer-wise distribution of spatial electric fields in the proposed NMS provides an explicit visualization of the underlying mechanism of EM modulation. In the transmission mode, the out-of-band suppression at the lower and upper regimes is dominated by two distinct spatial-field mechanisms. \textbf{Figure~\ref{fig:field_simu}a} reveals that at $f_{\text{TZ1}} = 8.2$~\si{GHz}, the spatial energy is strictly confined to the 1$^{\text{st}}$ layer, with the subsequent layers remaining unexcited, confirming a complete energy truncation at the incident boundary dictated by the intrinsic series resonance of the primary cascaded path. \textbf{Figure~\ref{fig:field_simu}c} depicts the multi-path cancellation mechanism at $f_{\text{TZ2}} = 11.6$~GHz that dictates the suppression of transmission. The incident field strongly excites the internal layers of the metasurface except the 5$^{\text{th}}$ layer. This phase cancellation between the primary cascaded path and the cross-coupling branch is consistent with the analysis in Section S1, where the cross-coupling predominantly influences the upper-sideband TZ. The two TZs can be adjusted by varying the structural parameters of the meta-atom, with parametric studies provided in SI section S5. \textbf{Figure~\ref{fig:field_simu}b} shows the resonant tunneling channel established at the center of the passband $f_0 = 10.1$~\si{GHz}, allowing the field to penetrate all stacked layers and maintain continuous forward energy propagation. By comparison, \textbf{Figure~\ref{fig:field_simu}d} captures the shielding mode, where the short circuit occurring at the second and fourth layers leads to a macroscopic impedance collapse of the structure. This results in a clamping of the propagating waves at the 1$^{\text{st}}$ layer and forms a broadband stopband.

\subsection{Experimental results under LIRF and HIRF}

\begin{figure}[H]
  \centering
  \includegraphics[width=\linewidth]{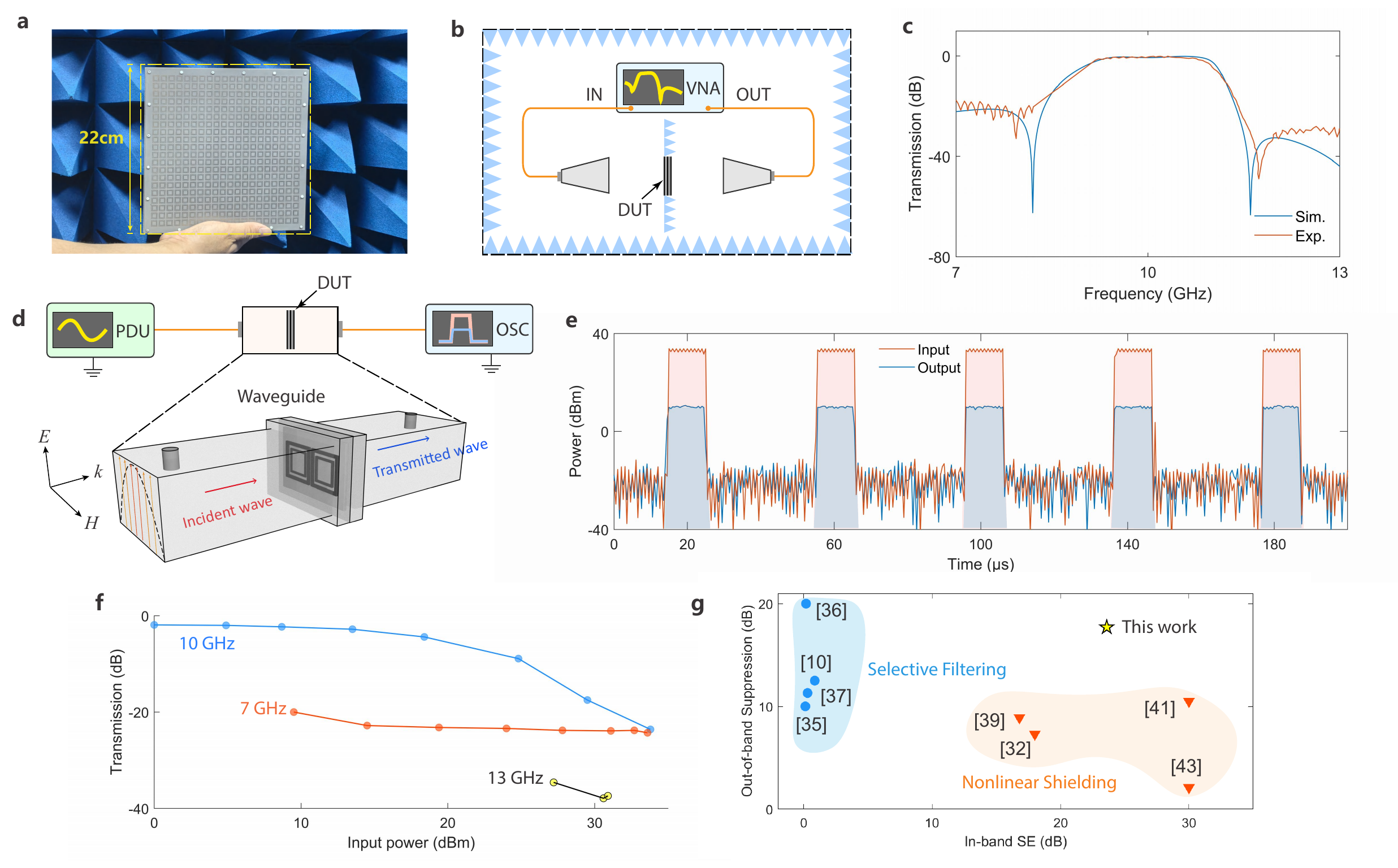}
  \caption{Experimental results of the proposed NMS under LIRF and HIRF. \textbf{a}. Photograph of the fabricated metasurface prototype. \textbf{b}. Free-space experimental setup in a microwave anechoic chamber for LIRF measurement. \textbf{c}. Measured and full-wave simulated free-space $S_{21}$ under LIRF. \textbf{d}. High-power waveguide measurement setup for HIRF evaluation. \textbf{e}. Input and output waveforms measured at 10~\si{GHz} under HIRF excitation in time domain. \textbf{f}. Power-dependent shielding effectiveness (SE) responses at representative in-band and out-of-band frequencies. \textbf{g}. Comparison of the proposed metasurface with reported studies in terms of out-of-band suppression and in-band SE under HIRF.}
  \label{fig:Experimental Demonstration}
\end{figure}

To validate the dual-state responses of the field-driven NMS, a prototype (20$\times$20 array, 220$\times$220 mm$^2$ footprint) was fabricated on top of F4BM220 substrate ($\varepsilon_{r} = 2.2$, $\tan \delta = 0.001$) via standard printed circuit board (PCB) technology, as shown in \textbf{Figure~\ref{fig:Experimental Demonstration}a}. The semiconductor diodes are integrated via surface-mount assembly on the 2$^{\text{nd}}$ and 4$^{\text{th}}$ layers, where the hollowed-out PMI foam frame is utilized to implement the air gap accommodating the packaging. We first test the transmission characteristics of the proposed NMS under LIRF. As shown in \textbf{Figure~\ref{fig:Experimental Demonstration}b}, the device under test (DUT) is placed in a standard microwave anechoic chamber and connected with a vector network analyzer (see \textbf{Methods} for more details). As expected, the measured free-space $|S_{21}|$ of the DUT agrees well with the full-wave simulated results, as plotted in \textbf{Figure~\ref{fig:Experimental Demonstration}c}. The passband maintains a low loss with IL $\leq$ 1~\si{dB} from 9.3 to 10.7~\si{GHz}, with out-of-band suppressions of 17.7 and 28.4~dB in the lower and upper stopbands, respectively. Around the transition region, despite a slight frequency shift attributed to the manufacturing tolerances, we still obtain steep roll-off rates of 20.6 and 31.8~\si{dB/GHz} at the lower and upper transition skirts, surpassing those reported in the literature, as compared in \textbf{Table~\ref{tab:comparison}}. We thus demonstrate a highly selective transmission together with anti-interference capability governed by the hybrid coupling topology in transmission mode.

We then evaluate the proposed NMS with respect to the field-driven adaptive EM protection under HIRF. As shown in \textbf{Figure~\ref{fig:Experimental Demonstration}d}, a waveguide injection platform has been built, in which a localized high-intensity RF is synthesized, relying on the spatial power confinement. More details about the experiment can be found in \textbf{Methods}. The time-domain input and output waveforms are captured at the passband center of 10~\si{GHz} to record the transient shielding behavior. As shown in \textbf{Figure~\ref{fig:Experimental Demonstration}e}, the maximal electric field in the waveguide center under HIRF can be calculated by the following formula:
\begin{equation}
E_{\text{TE10}} = \sqrt{\frac{4\eta P_{\text{TE10}}}{ab\sqrt{1 - (\lambda/2a)^2}}},
\label{eq.E}
\end{equation}
where $\eta$ represents the intrinsic wave impedance, $P_{\text{TE10}}$ is the input power, $\lambda$ is the free-space wavelength, and $a, b$ denote the broad and narrow dimensions of the waveguide, respectively. Considering a WR-90 rectangular waveguide with $a = 22.86$~mm and $b = 10.16$~mm, an input signal with a power of 33.8~dBm at 10~GHz produces a $E_{\text{TE10}}$ of 4.54~\si{kV/m} within the waveguide. This injection instantaneously triggers the DUT into the shielding mode, resulting in a measured attenuation of 23.6~\si{dB}. As shown in \textbf{Figure~\ref{fig:Experimental Demonstration}e}, the clamped output waveform without leading-edge overshoot indicates a transient protection against high-power spike leakage. 
Furthermore, we measure the power-dependent transmission of the proposed NMS at three typical frequencies, as shown in \textbf{Figure~\ref{fig:Experimental Demonstration}f}. For the 10~\si{GHz} incident signal, as the injected power increases over the recorded range, the transmission gradually transitions from the transparent state toward the shielding state. While for 7- and 13~\si{GHz}-incident signals, we cannot detect the output with low $P_{\text{TE10}}$ due to the attenuation falling below the noise floor of the spectrum analyzer. The thresholds of $P_{\text{TE10}}$ that make the output signals appear are about 9.5 and 27.2~\si{dBm} for 7 and 13~\si{GHz}, respectively. We also observe a lower transmission at 13~\si{GHz}, which is mainly attributed to a larger suppression in the upper stopband. The above HIRF measurements thus verify broadband shielding against the incident wave across the recorded spectrum. \textbf{Figure~\ref{fig:Experimental Demonstration}g} shows the comparison between our proposed NMS and other investigated metasurfaces from literature in terms of out-of-band suppression and in-band SE. For each work, the out-of-band suppression is represented by the weaker of the two stopbands. The numerical database is summarized as listed in \textbf{Table~\ref{tab:comparison}}. We observe an inherent trade-off in previous studies: selective filtering structures provide high spatial selectivity but lack high-power isolation, whereas conventional nonlinear shielding metasurfaces achieve the field-driven attenuation but sacrifice out-of-band selectivity. Consequently, our work physically integrates the hybrid coupling topology with field-driven impedance dynamics, achieving a single metasurface that performs dual functions in self-adaptive protection and anti-interference.   

\begin{table}[htbp]
\centering
\caption{Performance comparison between the proposed NMS and reported results from literature.}
\label{tab:comparison}
\footnotesize
\setlength{\tabcolsep}{3pt}
\begin{tabular}{@{}ccccccc@{}}
\hline
\multirow{2}{*}{Ref.} & \multirow{2}{*}{Mechanism} & Num.\ TZs\textsuperscript{1} & Roll-off rate\textsuperscript{2} (dB/GHz) & Out-of-band suppression (dB) & Measured & Measured \\
\cline{3-5}
 &  & Lower/Upper & Lower/Upper & Lower/Upper & IL/SE\textsuperscript{3} (dB) & Shielding BW\textsuperscript{4} (\%)\\
\hline
\cite{tian2025energy} & Loop coupling & 0\,/\,0 & \textasciitilde 12.1\,/\,12.1 & N/A\,/\,N/A & 2.4\,/\,20.2 & N/A  \\
\cite{hu2023design} & Parallel resonance & 0\,/\,0 & 12.3\,/\,N/A & 9.2{*}\,/\,5.3{*} & 0.6\,/\,16.8 & 29.41  \\
\cite{zhao2019power} & Parallel resonance & 0\,/\,0 & \textasciitilde 8.4\,/\,5.8 & 7.4{*}\,/\,7.3{*} & 1.2\,/\,18 & N/A  \\
\cite{yao2024high} & Parallel resonance & 0\,/\,0 & N/A\,/\,N/A & N/A\,/\,N/A & 0.17\,/\,21 & N/A \\
\cite{huang2024narrowband} & Dual resonance & 1\,/\,1 & 29.2\,/\,18.9 & 10.5\,/\,10.6 & 0.5\,/\,30 & 23.53  \\
\cite{tian2023wideband} & Multi-order resonance & 0\,/\,1 & 10\,/\,17.5 & N/A\,/\,35.4 & 1\,/\,24.2 & 20.69  \\
\cite{hu2021high} & Dual resonance & 1\,/\,1 & 8\,/\,2.4 & 2.1\,/\,10.8 & 1\,/\,30 & N/A \\
\textbf{This work} & \textbf{Hybrid coupling} & \textbf{2\,/\,2} & \textbf{20.6\,/\,31.8} & \textbf{17.7\,/\,28.4} & \textbf{0.1\,/\,23.6} & \textbf{60} \\
\hline
\end{tabular}

\vspace{1ex}
\raggedright
\textsuperscript{1}Num.\ TZs denotes the number of transmission zeros, excluding those at 0 and infinite frequencies;
\textsuperscript{2}Roll-off rate is defined as the attenuation from $-3$ to $-10$ dB divided by the corresponding transition bandwidth;
\textsuperscript{3}IL/SE: insertion loss / shielding effectiveness;
\textsuperscript{4}BW: bandwidth;
\textsuperscript{*}represents out-of-band suppression values estimated using an equal-bandwidth offset method: the passband bandwidth is taken as the $-3$ dB bandwidth of the transmissive state ($\text{BW}=f_{\text{upper}}-f_{\text{lower}}$), the anchor frequencies are offset from the passband edges by this bandwidth ($f_{\text{lower}}-\text{BW}$ lower and $f_{\text{upper}}+\text{BW}$ upper), and suppression is read as $|S_{21}|$ at these anchor frequencies;
\textasciitilde\ represents estimated values.
\end{table}

The elevated roll-off rates (20.6 and 31.8~\si{dB/GHz}) of the proposed NMS sharply confine the transmission window, hence enabling effective suppression of out-of-band interference even at closely neighboring bands. According to the built hybrid coupling framework, the spectral characteristics, including TZ frequency, roll-off rate and the center frequency of passband, can be modulated with the geometrical size. More details about parameter scanning can be found in SI Section S5. In addition, the designed metasurface maintains angular robustness up to $45^\circ$ oblique incidence, as analyzed in SI Section S3. In the transmission mode, the IL remains below 3~\si{dB}, while in the shielding mode the SE within the original passband stays above 20.4~dB for both TE and TM polarizations. Beyond this angular stability, the obtained 60\% shielding bandwidth under HIRF is much higher than the reported values around 20--30\% (see \textbf{Table~\ref{tab:comparison}}). This effectively eliminates the spectral leakage windows that could compromise receiver frontends. Among prior works combining selectivity with nonlinear shielding, the quasi-elliptic design in Ref.~\cite{tian2023wideband} provides only a single-sided stopband, while Ref.~\cite{huang2024narrowband} achieves dual-sided selectivity but only with an out-of-band suppression of 10~\si{dB} and a narrow measured shielding bandwidth around 23.5~$\%$. Despite the above advances, the designed five-layer stacked configuration introduces fabrication complexity. Simplifying the structure while preserving the hybrid coupling degrees of freedom remains an open challenge. Moreover, although our design achieves higher out-of-band suppression relative to existing NMSs, its performance still lags behind that of state-of-the-art frequency-selective surfaces, revealing a clear pathway for future enhancement.

Looking ahead, the proposed design methodology, which integrates hybrid coupling topology with field-driven nonlinearity, offers a viable path towards higher-frequency operation. Extending this architecture to the Ku‑ or Ka‑band will require proportionate downscaling of the meta‑atom dimensions, along with the implementation of microfabrication or thin‑film deposition techniques to ensure the structural integrity required for the designated cross‑coupling paths. Beyond frequency scaling, the present framework is readily compatible with alternative nonlinear components, including varactors and phase‑change materials, enabling either an adjustable switching threshold or continuous electrical tuning between the transmissive and shielding states. Furthermore, the potential for integration with conformable or flexible substrates could significantly expand the deployment space of this approach in adaptive electromagnetic protection scenarios.

\section{Conclusions}

We demonstrated a field-driven nonlinear metasurface that performs a self-adaptive transition between high-selectivity transmission and broadband shielding. The geometrical size of this structure is fixed by optimizing a hybrid coupling topology. From measurement, we obtain a highly selective passband with roll-off rate larger than 20.6~\si{dB/GHz} in transmission mode and a shielding with 60~$\%$ bandwidth. The measured $S_{21}$ curves agree well with theoretical analysis and simulation, allowing us to achieve an effective protection against both out-of-band EM interference and in-band high-intensity radiation. The extracted performance indicators, including roll-off rate, the operating bandwidths, out-of-band suppression and shielding effectiveness, are enhanced a lot compared with the previously reported results. This work not only advances the fundamental understanding of multilayer hybrid coupling in wave tuning but also potentially enables studies into the use of nonlinear metasurfaces for ensuring the spatial EM safety of electronic information systems.

\section{Methods}

\threesubsection{Simulations}\\
All 3D electromagnetic simulations are conducted using CST Studio Suite. To evaluate the $|S_{21}|$ of the proposed NMS, full-wave simulations are performed with periodic boundary conditions and Floquet port excitations applied to the meta-atom, emulating a free-space infinite array. The CST field-circuit co-simulator is employed to link the 3D full-wave electromagnetic boundaries with the lumped circuit solver. The setup comprises a transmitting Port 1 and a receiving Port 2 separated by 110 mm, with the meta-atom placed in the middle. Discrete ports are defined across the structural gaps at the specific locations where the diodes are loaded, with the corresponding diode SPICE models assigned to them. Driven by a sinusoidal excitation with a 1.5 ns rise time, the field-dependent transmission characteristics are evaluated by scaling the electric field intensity and tuning the frequency of the incident wave.

\threesubsection{Experiments}\\
As illustrated in Figure S3a, the LIRF free-space measurement was conducted in a standard microwave anechoic chamber. The fabricated prototype was mouanted within a central aperture of an absorbing screen. Two broadband horn antennas were symmetrically positioned on both sides of the sample to illuminate and receive plane waves under normal incidence. Both antennas were connected to a vector network analyzer (Keysight N5227A) to record the spatial $|S_{21}|$. As described in SI section S4, the HIRF evaluation was conducted utilizing a high-power waveguide measurement setup. The microwave signal is synthesized by a vector signal generator (VSG) and subsequently amplified by a power amplifier (PA). Crucially, to protect the PA from the severe reflected energy induced by the broadband impedance collapse of the metasurface in State II, a circulator terminated with a matched load is integrated into the main transmission path. The amplified intense field is then injected into a standard WR-90 rectangular waveguide, which encapsulates a $2 \times 1$ meta-atom array. At the receiving end, the transmitted signal is captured by a spectrum analyzer, which is safeguarded by a high-power attenuator to prevent frontend burnout. To extract the actual $|S_{21}|$, an empty waveguide is first measured to establish a calibration baseline. The sample is then inserted, and the precise $|S_{21}|$ is determined by normalizing the received power of the loaded sample against the empty waveguide baseline. During the evaluation, the field intensity and operating frequency of the injected wave are systematically swept by controlling the output power and frequency of the VSG.

\medskip
\textbf{Supporting Information} \par 
Supporting Information is available from the Wiley Online Library or from the author.

\medskip
\textbf{Acknowledgements}
\par
This work was supported by the National Natural Science Foundation of China under Grant No. 62401592 and the Young Scientists Innovation Fund Project of NUDT under Grant No. ZK24-10.

\medskip

\bibliographystyle{MSP}
\bibliography{MSP-template}

\clearpage

\section*{\centering Supporting Information}
\addcontentsline{toc}{section}{Supporting Information}

\setcounter{subsection}{0}
\setcounter{figure}{0}
\setcounter{table}{0}
\setcounter{equation}{0}
\renewcommand{\thesubsection}{S\arabic{subsection}}
\renewcommand{\thefigure}{S\arabic{figure}}
\renewcommand{\thetable}{S\arabic{table}}
\renewcommand{\theequation}{S\arabic{equation}}

\begin{center}
    \vspace*{1em}
    {\Large \textbf{Field-driven nonlinear metasurface: self-adaptive transitions between high-selectivity transmission and broadband shielding}} \\[1em]
     {\large Yuan Xu$^{1,3}$, Hanqing Liu$^{*,1,3}$, Shangjing Xi$^{1}$, Xiaodi Zhang$^{1}$, Yanqing Cheng$^{1}$, Wenye Ji$^{*,2}$, and Peiguo Liu$^{*,1}$}
\end{center}
\vspace{2em}


\subsection{Equivalent Circuit Model Analysis}
Given the equivalent circuit model (ECM) established in the main text, this section theoretically elucidates the generation mechanism of the transmission zeros (TZs) required for achieving high-selectivity transmission, and subsequently presents the extracted lumped parameters that fulfill the targeted dual-state response consisting of the high-selectivity transmission mode and the broadband shielding mode.

The analysis begins with the definition of the generalized Chebyshev filtering function \cite{cameron2002general}. The transmission coefficient $S_{21}(s)$ is expressed as the ratio of two characteristic polynomials:
\begin{equation}
\label{eq:S21_poly}
S_{21}(s) = \frac{P(s)}{\varepsilon E(s)}
\end{equation}
where $s = j\omega$ is the complex frequency variable, and $\varepsilon$ is a constant normalizing the equiripple level. The roots of the polynomial $P(s)$ correspond to the TZs of the network, while the roots of $E(s)$ represent the poles. To realize the finite-frequency TZs required for high out-of-band suppression, $P(s)$ must possess finite roots, which mathematically dictates the use of a non-cascaded coupling topology. Consequently, the network topology in the diode-OFF state is mapped to a generalized $(N+2)$ coupling matrix $[M]$ \cite{ESScameron2003advanced}:
\begin{equation}
\label{eq:coupling_matrix}
[M] =
\begin{bmatrix}
0 & M_{S1} & 0 & 0 & 0 \\
M_{S1} & M_{11} & M_{12} & M_{13}(\omega) & 0 \\
0 & M_{12} & M_{22} & M_{23} & 0 \\
0 & M_{13}(\omega) & M_{23} & M_{33} & M_{3L} \\
0 & 0 & 0 & M_{3L} & 0
\end{bmatrix}
\end{equation}
Within this generalized coupling matrix $[M]$, the diagonal entries $M_{ii}$ represent the self-couplings, while the off-diagonal entries $M_{ij}$ denote the inter-resonator couplings. The zero entries (e.g., $M_{S2}=0$) reflect the physical constraints of the cascaded topology, where no direct coupling exists between non-adjacent nodes. Crucially, a frequency-variant cross-coupling element $M_{13}(\omega)$ is incorporated to generate the TZs \cite{tamiazzo2016synthesis}. In the physical ECM, this is actualized by the cross-coupling branch comprising $C_{m13}$ and $L_{m13}$ (as defined in the ECM). The corresponding frequency-variant cross-coupling coefficient $M_{13}(\omega)$ is derived as:
\begin{equation}
\label{eq:mapping}
M_{13}(\omega) \propto \omega C_{m13} - \frac{1}{\omega L_{m13}}
\end{equation}

Equation~(\ref{eq:mapping}) forms the physical basis for the multipath signal cancellation mechanism. A TZ is generated when the signal propagating through the main path ($1 \to 2 \to 3$) destructively interferes with the signal through the cross-coupling path ($1 \to 3$). This condition is mathematically expressed as:
\begin{equation}
\label{eq:cancellation}
\frac{M_{12} M_{23}}{\Omega - M_{22}} + M_{13}(\Omega) = 0
\end{equation}
This cancellation mechanism governs the TZ placement of the present metasurface.

In this study, the targeted high-selectivity transmission mode is centered near the middle of the X-band with a fractional bandwidth (FBW) of approximately 15\%, while simultaneously satisfying the broadband impedance collapse criteria in the shielding mode. The performance of the equivalent circuit was simulated and optimized using the commercial software Advanced Design System (ADS). The final extracted circuit component values are summarized in Table~\ref{tab:circuit_params}.

\begin{table}[H]
\centering
\caption{Detailed Circuit Parameters of the Proposed ESS}
\label{tab:circuit_params}
\scriptsize
\setlength{\tabcolsep}{3.5pt} 
\begin{tabular}{@{}lcccccc@{}}
\hline
\textbf{Param.} & $h_1$ & $h_2$ & $h_3$ & $C_{1}$ & $L_{1}$ & $C'_{1}$ \\
\textbf{Value}  & 4 mm & 1 mm & 0.254 mm & 21.5 fF & 17.46 nH & 12.23 fF \\
\hline
\textbf{Param.} & $L'_{1}$ & $C_{2}$ & $L_{2}$ & $C_{3}$ & $L_{3}$ & $C'_{3}$ \\
\textbf{Value}  & 15.8 nH & 0.712 pF & 45 pH & 48 fF & 19.4 nH & 12.44 fF \\
\hline
\textbf{Param.} & $L'_{3}$ & $C_{m13}$ & $L_{m13}$ & $C_{off}$ & $R_{on}$ & \\
\textbf{Value}  & 8.6 nH & 0.25 fF & 0.33 nH & 30 fF & 4.2 $\Omega$ & \\
\hline
\end{tabular}
\end{table}

\begin{figure}
  \includegraphics[width=\linewidth]{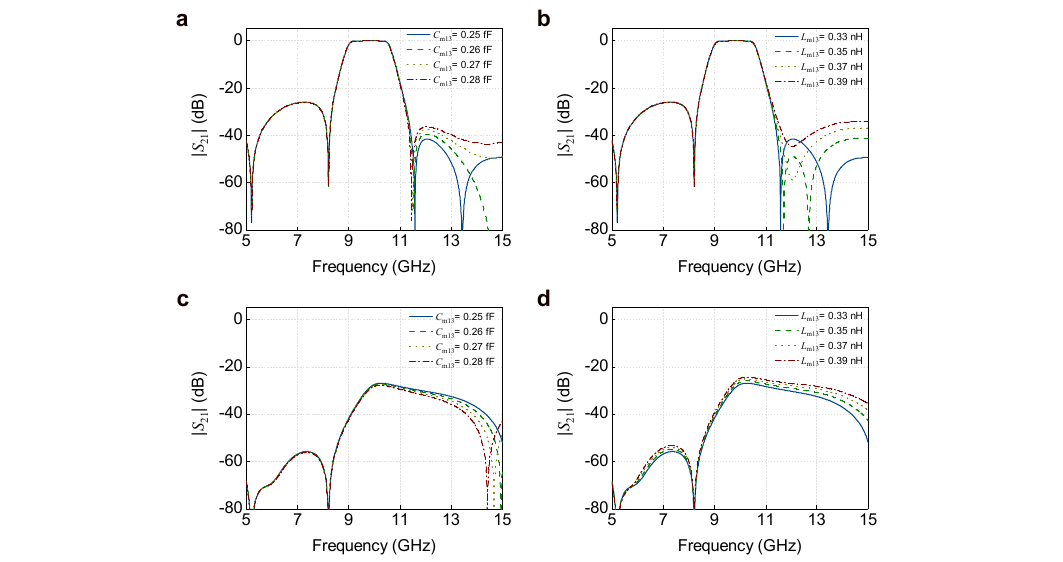}
  \caption{Parametric analysis of the cross-coupling elements on the transmission coefficient ($|S_{21}|$) under different operating states. Variation of a) $C_{m13}$ and b) $L_{m13}$ in the transmission mode. Variation of c) $C_{m13}$ and d) $L_{m13}$ in the shielding mode.}
  \label{fig:parametric_sweep}
\end{figure}

To further investigate the impact of the cross-coupling, simulations are conducted for the elements ($C_{m13}$ and $L_{m13}$). As depicted in Figures~\ref{fig:parametric_sweep}a and \ref{fig:parametric_sweep}b, in the transmission mode, these elements primarily govern the locations of the TZs and the out-of-band suppression depth in the upper sideband, thereby influencing the filtering response at the adjacent upper band and higher frequencies, while the lower-sideband TZs are determined by the primary cascaded path. Furthermore, under the shielding mode as shown in Figures~\ref{fig:parametric_sweep}c and \ref{fig:parametric_sweep}d, variations in $C_{m13}$ and $L_{m13}$ affect the shielding effectiveness (SE). Specifically, increasing $C_{m13}$ while decreasing $L_{m13}$ enhances the SE, whereas decreasing $C_{m13}$ while increasing $L_{m13}$ weakens it.

\subsection{Relationship Between Equivalent Circuit Elements and Structural Parameters}

For the individual metallic layers, the continuous conductive traces and isolating gaps function as lumped inductors and capacitors, respectively. Based on the impedance calculation models for metallic strip-gap structures \cite{sung2006frequency,yao2020study}, the approximate $L$ and $C$ models of the meta-atom can be obtained as follows:

$$
L \approx \frac{Z_{0} p \sqrt{\epsilon_{r}\mu_{r}}}{2\pi} \left[ \ln\left(\csc\left(\frac{\pi w}{2p}\right)\right) + G(p, w, \lambda) \right]
$$

$$
C \approx 4\frac{p\sqrt{\epsilon_{r}\mu_{r}}}{2\pi Z_{0}}\left[ \ln\left(\csc\left(\frac{\pi g}{2p}\right)\right) + G(p, g, \lambda) \right]\epsilon_{eff}
$$

where $p$, $w$, and $g$ denote the unit-cell periodicity, trace width, and gap width, respectively. The term $G(p, x, \lambda)$, with $x$ representing either $w$ or $g$, is defined as:

$$
G(p, x, \lambda) = \frac{0.5(1-\beta^2)^2 \left[ 2A\left(1-\frac{\beta^2}{4}\right) + 4\beta^2 A^2 \right]}{\left(1-\frac{\beta^2}{4}\right) + 2A\beta^2\left(1+\frac{\beta^2}{2}-\frac{\beta^4}{8}\right) + 2\beta^6 A^2}
$$

with $A = 1/\sqrt{1-(p/\lambda)^2} - 1$ and $\beta = \sin\left(\frac{\pi x}{2p}\right)$. Since this analytical model is applicable under the sub-wavelength condition ($p \ll \lambda$), which is satisfied by the designed meta-atom, and uses the optimized ECM values in Table~S1 as target values, it provides the initial structural parameters for each layer, followed by optimization in CST Microwave Studio to obtain the multilayer metasurface reported in the main text.

\subsection{Angular Stability Analysis}

\begin{figure}[H]
  \includegraphics[width=\linewidth]{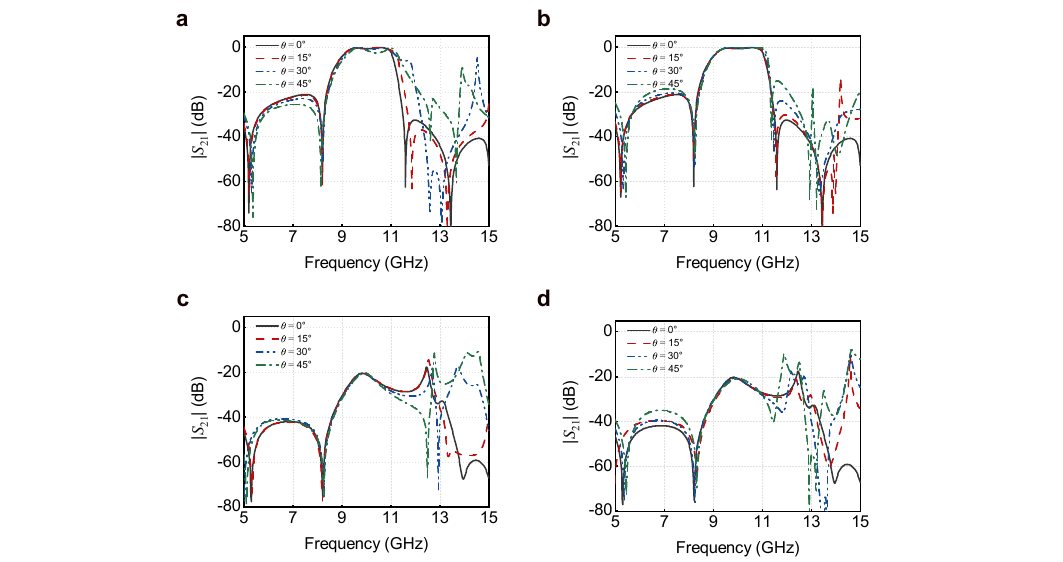}
  \caption{Simulated $|S_{21}|$ of the proposed metasurface under oblique incidence ($\theta = 0^{\circ}$ to $45^{\circ}$). a) TE and b) TM polarizations in the transmission state. c) TE and d) TM polarizations in the shielding state.}
  \label{fig:angle_sweep}
\end{figure}

The angular stability of the designed metasurface is characterized under oblique incidence angles ($\theta$) up to $45^{\circ}$ for both TE and TM polarizations, as depicted in Figure~\ref{fig:angle_sweep}. In the transmission state, as the incident angle increases, the TE polarization in Figure~\ref{fig:angle_sweep}a exhibits a passband shift toward higher frequencies along with TZ drifts induced by spatial dispersion. Nevertheless, the maximum passband insertion loss (IL) remains below 2.54 dB. For the TM polarization in Figure~\ref{fig:angle_sweep}b, the passband remains stable up to $45^{\circ}$. Although the TZs also experience frequency deviations, an effective out-of-band suppression is maintained to guarantee the filtering response. Overall, the designed spatial filtering profile exhibits angular stability under the transmission state. 

In the shielding state shown in Figure~\ref{fig:angle_sweep}c and d, although the higher-frequency part of the response exhibits frequency shifts as the oblique angle increases up to $45^{\circ}$, the SE within the original passband remains above 20.4~dB for both polarizations, while the lower-frequency regime shows negligible variation. Consequently, despite the high-frequency deviations, the metasurface preserves reliable dual-state capability under oblique incidence.

\subsection{Photographs of the Experimental Setups}

\begin{figure}[H]
  \includegraphics[width=\linewidth]{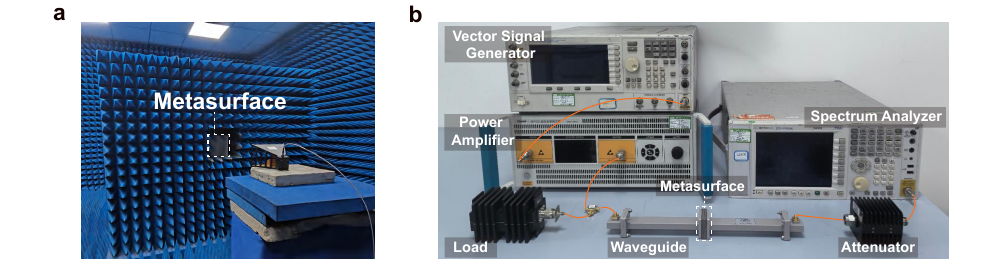}
  \caption{Photographs of the experimental setups. a) Free-space measurement setup in a microwave anechoic chamber for low-intensity radiation field evaluation. b) High-power waveguide measurement setup for high-intensity radiation field evaluation.}
  \label{fig:setups}
\end{figure}

Photographs of the experimental setups for both low-power free-space and high-power waveguide measurements are presented in Figure~\ref{fig:setups}.

\subsection{Structural Parameter Analysis in Full-Wave Simulation}

\begin{figure}[H]
  \includegraphics[width=\linewidth]{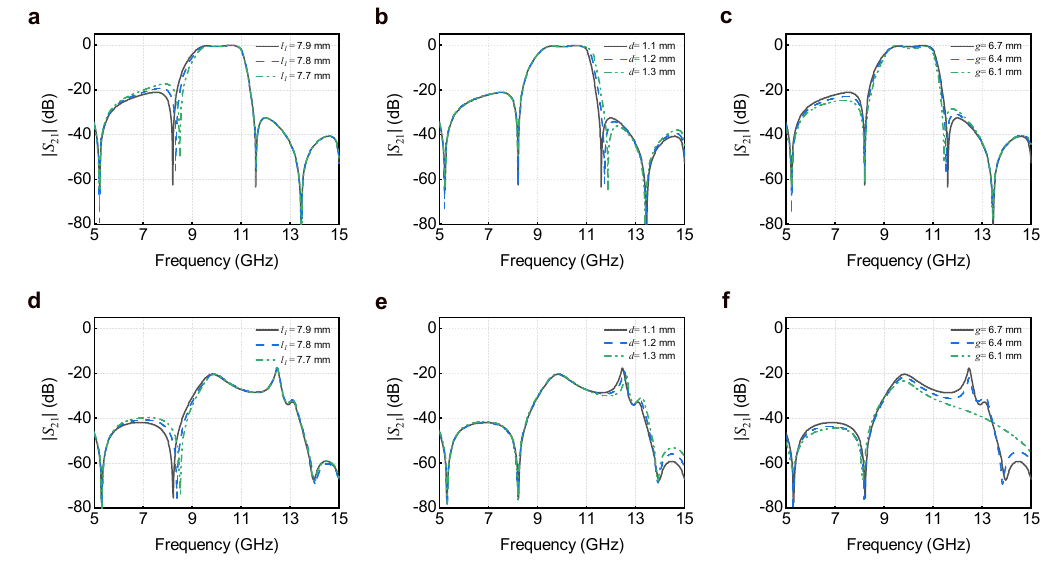}
  \caption{Structural parameter analysis of $l_1$, $d$, and $g$ on the transmission coefficient $|S_{21}|$. a-c) Transmission mode. d-f) Shielding mode. Each column from left to right corresponds to the variation of $l_1$, $d$, and $g$.}
  \label{fig:Full-wave analysis}
\end{figure}

Full-wave parametric sweeps are carried out for three representative structural parameters: $l_1$, $d$, and $g$. The corresponding transmission coefficients under both the transmission and shielding modes are shown in Figure~\ref{fig:Full-wave analysis}. These results illustrate that the spectral characteristics can be adjusted, for instance, in terms of transmission-zero locations, roll-off rates, passband bandwidth, and shielding performance.

As shown in Figure~\ref{fig:Full-wave analysis}a and d, the parameter $l_1$ primarily tunes the lower-sideband TZ closest to the passband. When $l_1$ decreases, the outer-frame inductance on the first layer is reduced, and this TZ shifts toward higher frequencies. As a result, the lower transition skirt becomes steeper and the passband bandwidth is correspondingly narrowed. The same frequency shift can also be observed for the corresponding spectral feature in the shielding mode. In contrast, as shown in Figure~\ref{fig:Full-wave analysis}b and e, the parameter $d$ exhibits the strongest sensitivity to the upper-sideband TZ closest to the passband in the full-wave model. Increasing $d$ reduces the effective capacitance associated with the second-layer gap, thereby shifting this TZ toward higher frequencies. This shift relaxes the upper transition skirt and slightly broadens the passband, mainly on the high-frequency side. Finally, Figure~\ref{fig:Full-wave analysis}c and f show that the parameter $g$ mainly controls the overall transition steepness and thus the roll-off rates on both sides. Decreasing $g$ sharpens both transition skirts, increases the roll-off rates, and tends to enhance the SE in the shielding mode, while also increasing the IL within the passband. Overall, these results show that the proposed metasurface provides multiple geometric tuning knobs for flexible spectral reconfiguration.


\end{document}